\documentclass[aps,prl,twocolumn,showpacs,amsmath,amssymb]{revtex4}
\usepackage{graphicx}% Include figure files
\usepackage{dcolumn}% Align table columns on decimal point
\usepackage{bm}% bold math

\begin{document}

\title{Temperature dependence of the resistance of a phase-slip line in a thin
superconducting film}

\author{E.~V.~Il'ichev}
\altaffiliation [ Now at ] { Institute for Physical High
Technology, P.O. Box 100239, D-07702 Jena, Germany}
\author{V.~I.~Kuznetsov}
\email[ Electronic address: ] {kvi@ipmt-hpm.ac.ru}
\author{V.~A.~Tulin}
\affiliation{Institute of Microelectronics Technology and High
Purity Materials, Russian Academy of Sciences, 142432
Chernogolovka, Moskow Region, Russia}

\date{\today}

\begin{abstract}
An experimental investigation was made of the temperature
dependence of the first step of a phase-slip line in a thin
superconducting tin film. The depth of penetration of a
nonequilibrium longitudinal electric field into the superconductor
was determined near the critical temperature. A comparison was
made with theoretical investigations of one-dimensional structures
containing phase-slip centers. The experimental results were found
to be in good agreement with the theory when the mechanism of
mixing of electron-like and hole-like branches of the
quasiparticle spectrum was governed by the elastic scattering of
the excitations. This was one more experimental confirmation that
a phase-slip line is a two-dimensional analog of a phase-slip
center.
\end{abstract}

% insert suggested PACS numbers in braces on next line
% insert suggested keywords - APS authors don't need to do this
%\keywords{}

%\maketitle must follow title, authors, abstract, \pacs, and \keywords
\maketitle

The passage of a constant transport current exceeding the critical
value gives rise to a spatially inhomogeneous resistive state in a
superconducting film. A special feature of this state is the
existence of an energy gap in the spectrum of quasiparticle
excitations, i.e., the existence of a superconducting current, and
also of a normal dissipative current giving rise to a voltage
drop. A narrow film of width $W$ less than the coherence length
$\xi (T)$ splits into localized resistive regions known as the
phase-slip centers \cite{i}. Phase-slip lines \cite{ii} may form
in a wide film characterized by $W > \xi (T)$ if the heat removal
conditions are good.

The depth of penetration $l_{E}$ of a nonequilibrium longitudinal
electric field determines the resistance $R_{0}$ to the flow of
the current of a single phase-slip center \cite{i} and a
phase-slip line \cite{ii}, where
\begin{eqnarray} R_{0} = 2 \rho_{n}l_{E}/Wd \;,
\label{eq:1}
\end{eqnarray}
where $\rho_{n}$ is the resistivity in the normal state and $d$ is
the film thickness. In a more or less homogeneous film the above
expression can be rewritten in the form
\begin{eqnarray} R_{0}=2R_{n}l_{E} /L \;,
\label{eq:2}
\end{eqnarray}
where R$_{n}$ is the resistance in the normal state and L is the
length of the whole film.

The temperature dependence $R_{0}(T)$ can be used to judge the
temperature dependence of $l_{E}$. Information on the depth of
penetration of the electric field can be obtained also directly by
microprobe located near a single phase-slip center \cite{iii} and
a phase-slip line \cite{ii}. The depth of penetration of the
electric field is governed by various mechanisms of relaxation of
the difference between the populations of electron- and hole-like
branches of the energy spectrum of a superconductor
\cite{iiii,iiiii}.
 They include an inelastic mechanism of electron -phonon collisions, an
"elastic" mechanism in the case of a sufficiently strong current
$j_{s}$ of the condensate, and other "elastic" mechanisms
involving the scattering by paramagnetic impurities and the
anisotropy and inhomogeneity of the energy gap. Different
mechanisms have different temperature dependence. An experimental
investigation of the dependence $l_{E}(T)$ can be used to
determine which mechanism is active in a given situation.

As demonstrated in many theoretical and experimental
investigations \cite{i,iii}, an inhomogeneous distribution of the
longitudinal field in a phase-slip center is formed primarily by
the inelastic electron-phonon scattering processes. In this case
the depth of penetration of an electric field into a
superconductor with an energy gap is \cite{iiiii}
\begin{eqnarray} l^{n}_{E}= \sqrt{D\tau_{Q}} = \sqrt{D\tau_{\varepsilon}\frac{4kT}{\pi \Delta}} \;,
\label{eq:3}
\end{eqnarray}
where $\Delta$ is the energy gap, $D=(1/3)v_{F}l$ is the electron
diffusion coefficient ($v_{F}$ is the Fermi velocity and $l$ is
the mean free path), $\tau_{Q}$ is the relaxation time of the
asymmetry of the populations of the branches of the quasiparticle
spectrum, and $\tau_{\varepsilon }$ is the relaxation time. Near
$T_{c}$ in the case of the inelastic mechanism we have
\begin{eqnarray}l^{n}_{E}=A(D\tau_{\varepsilon})^{1/2}(1-\tau)^{-1/4}\;,
\label{eq:4}
\end{eqnarray}
where $A$ is the numerical coefficient of the order of unity,
$\tau=T/T_{c}$, and $T_{c}$ is the critical temperature.

It is shown in Refs. \cite{iiii,iiiiii,iiiiiii,iiiiiiii} that
another mechanism of penetration of an electric field is the
elastic scattering of excitations resulting in mixing of the
electron and hole branches of the spectrum. It is important if we
allow for the finite velocity of flow of the superconducting
condensate. At temperatures somewhat further from $T_{c}$ these
processes may predominate over the inelastic mechanism. The depth
of penetration of an electric field in the case of the elastic
scattering mechanism is \cite{iiiiii,iiiiiii}
\begin{eqnarray} l^{el}_{E}= \frac{\pi \varepsilon }{ P_{s} \Delta \sqrt{2}}  \;,
 \label{eq:5}
\end{eqnarray}
where $P_{s}$ is the superfluid momentum and $\varepsilon$ is a
characteristic energy. Near $T_{c}$, we obtain
\begin{eqnarray} l^{el}_{E}=B(\xi_{0}l)^{1/2}(1-\tau)^{-1} \;,
\label{eq:6}
\end{eqnarray}
where $B$ is a numerical coefficient of the order of unity and
$\xi_{0}$ is the coherence length of a pure superconductor.

The limiting conditions for the realization of the elastic and
inelastic case are found theoretically in Ref. \cite{iiii}. The
inelastic mixing mechanisms can be neglected if
\begin{eqnarray} l^{el}_{E}<<l^{n}_{E } \;.
\label{eq:7}
\end{eqnarray}
In the case of a homogeneous sample, we have
\begin{eqnarray} (1-\tau)>> \left [ \frac{\hbar}{\tau_{\varepsilon}kT_c} \right ] ^{2/3} \;.
\label{eq:8}
\end{eqnarray}
It follows from the above inequality that when the temperature
shifts somewhat from $T_{c}$, the elastic processes of mixing of
electron and hole branches of the quasiparticle spectrum may
predominate in such a homogeneous sample. Moreover, the condition
(8) depends on the energy relaxation time $\tau_{\varepsilon}$.
Theoretical estimates of $\tau_{\varepsilon}$ are given in Ref.
\cite{iiiiiiiii}, but the experimental values of
$\tau_{\varepsilon}$ are subject to a large scatter. Therefore, at
a given temperature we may have the same mixing mechanism in
different samples.

Phase-slip centers have been investigated quite thoroughly both
theoretically and experimentally \cite{i, iiiiiiiii}, but there
are practically no theoretical studies of phase-slip lines. These
lines have the following properties: 1) motion of vortices until
the first phase-slip line is formed; 2) creation of a magnetic
field by the transport current (in the case of phase-slip centers
this field is ignored because of the small width of the films).
This makes it difficult to analyze phase-slip lines theoretically.
The temperature dependence of the resistance of these lines has
not yet been investigated sufficiently thoroughly. Even in the
case of phase-slip centers the early treatments of Tinkham et al.
\cite{iiiiiiiii} failed to detect a temperature dependence of the
differential resistance of these centers. This dependence was
however observed subsequently \cite{iiiiiiiiii, iiiiiiiiiii}. In
the case of wide films the temperature dependence of the
differential resistance of a phase-slip line was first
investigated by us \cite{iiiiiiiiiiii} and we found that the depth
of penetration of an electric field is governed by the inelastic
scattering mechanism.

Variation of the initial data of a sample allows us to consider
the elastic mechanism. The temperature dependence of the
resistance of a phase-slip line under the elastic scattering
conditions is the subject to the present paper. We prepared
samples with a much lower resistivity $\rho_{n} =8 \ast 10^{-7}
 \Omega \bullet$ cm than used by us in Ref. \cite{iiiiiiiiiiii},
which made it possible to widen the temperature range of the
existence of phase-slip lines until they were transformed into
resistive domains, which increased the probability of a
realization of the elastic case. Moreover, in the case of the new
samples the time $\tau_{\varepsilon }$ was several times longer
than in the study reported in Ref. \cite{iiiiiiiiiiii}.

Our samples were tin films $d \approx 2000$ {\AA} thick, $W = 70$
$\mu$m wide, $L=2$ mm long and $R_{n} \approx 1.14$   $\Omega.$
They were formed by thermal evaporation on silicon substrates kept
at room temperature. Weak spots (cuts) at side boundaries of a
film strip were formed to facilitate the appearance of isolated
phase-slip lines. The current-voltage characteristic was recorded
at different temperatures of a helium bath near $T_{c} \approx
3.91$ K. This characteristic was stepped and it consisted of a
series of linear regions with a differential resistance which was
a multiple of the resistance $R_{0}$ of a single phase-slip line,
similar to that reported in Ref. \cite{iiiiiiiiiiii}.

The temperature dependence of the resistance of the first
phase-slip line (Fig. 1) was qualitatively similar to the
temperature dependence reported by us earlier \cite{iiiiiiiiiiii}.
However, in the present case, single phase-slip lines were
manifested less clearly in the current-voltage characteristic and
merging of several phase-slip lines usually occurred. The
appearance of the first step due to a single phase-slip line near
the critical current and critical temperature was observed for
samples with a strong inhomogeneous boundary (the dimensions of
the inhomogeneity were less than the depth of penetration of an
electric field). The main difference was that the resistance of a
phase-slip line had a stronger temperature dependence (Fig. 1) in
the isothermal interval, which was  $ \approx$ $3.86 \div 3.90$ \
K in our case.

\begin{figure}
\includegraphics[width = 1.0\linewidth]{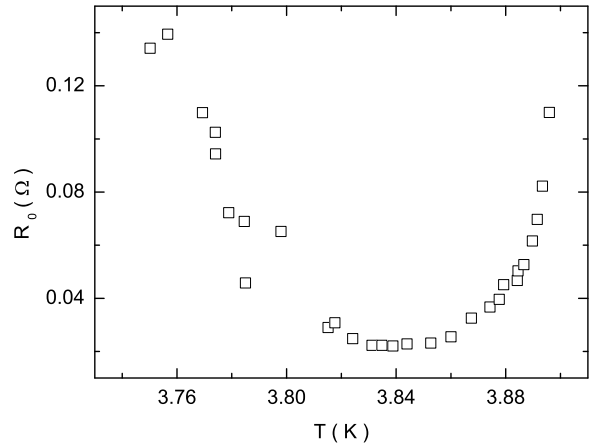}
\caption{\label{fig. 1} Temperature dependence of the resistance
of a phase-slip line. The increase in the resistance as a result
of cooling in the temperature interval less than $T \simeq 3.84$ K
was due to conversion of this phase-slip line into a elementary
resistive domain because of the deviation from isothermal
conditions.}
\end{figure}

\begin{figure}
\includegraphics[width = 1.0\linewidth]{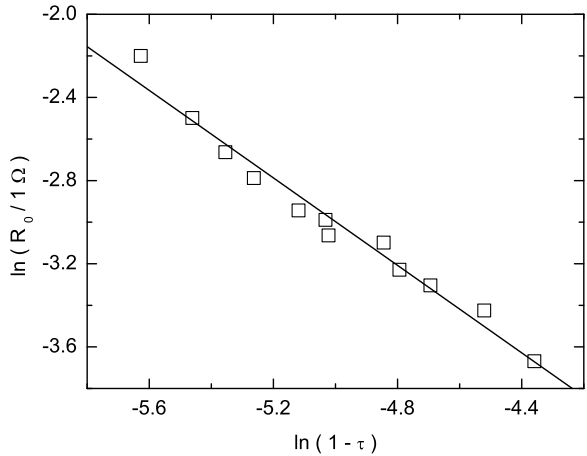}
\caption{\label{fig. 2}Temperature dependence of the resistance of
a phase-slip line in the isothermal interval. }
\end{figure}

The results of a quantitative analysis obtained for one of our
samples are presented in Fig. 2. The temperature dependence of the
dimensionless resistance $R_{0}/1 \Omega$ on $( 1-\tau )$ in the
isothermal interval was plotted on a logarithmic scale. The
experimental points fitted well a straight line with the slope
$\approx -1$ and the continuation (extrapolation) of this line
intersected the ordinate at $ \ln M_{exp} \approx -8.26$. Hence,
we found the experimental value $M_{exp} \approx 2.59 \ast
10^{-4}$ $\Omega$. Assuming that
\begin{eqnarray} l^{el}_{E}=B(\xi_{0}l)^{1/2}(1-\tau)^{-n} \;,
\label{eq:9}
\end{eqnarray}
we obtained experimental value of the resistance
\begin{eqnarray} R_{0}^{ exp }=R_{n} \frac{2l_{E}}{L}=M(1-\tau)^{-n} \;,
\label{eq:10}
\end{eqnarray}
where \begin{eqnarray} M = \frac{2R_{n}B(\xi_{0}l)^{1/2}}{L} \;,
\label{eq:11}
\end{eqnarray}
$\xi_{0} =  2.3 \ast 10^{-5}$ cm is the coherence length of tin
and $l=2 \ast 10^{-5}$ cm is the mean free path in the tin film
obtained from $\rho_{n}l \simeq 1.6 \ast 10^{-11}\Omega \bullet
cm^{2}$ \cite{iiiiiiiiiiii}.

A comparison of the experimental value of $M_{exp}$, deduced from
Fig. 2, with $M \simeq B (2.44 \ast 10^{-4})$ $\Omega,$ deduced
from (11) indicated that the numerical coefficient was $B \simeq
1$. Therefore, the depth of penetration of an electric field found
from the experiment results was given by Eq. (9), where $B \simeq
1$ and $n \simeq 1$. This was in agreement with the theoretical
predictions \cite{iiiiiii} and with the experimental results
\cite{iiiiiiiiii} obtained for narrow films in which phase-slip
centers formed. In the elastic case characterized by $l_{E}^{n} >
l_{E}^{el}$ we could estimate the energy relaxation time
$\tau_{\varepsilon}$. This time was of the order of $10^{-9}$ s,
which was close to the value found in Ref. \cite{iiiiiiiiii}, when
both (elastic and inelastic) scattering mechanisms were active
under the conditions of formation of phase-slip centers. This
value of $\tau_{\varepsilon }$ was several times greater than that
used in Ref. \cite{iiiiiiiiiiii}, where the inelastic mechanism
was observed.

We thus demonstrated for the first time that in the case of wide
films containing phase-slip lines, when the transport current was
slightly higher than the critical value, the elastic relaxation
mechanism of the imbalance of populations of the electron- and
hole-like branches of the quasiparticle spectrum of excitations
could predominate and govern the depth of penetration of an
electric field and, consequently, the temperature dependence of
the resistance of a phase-slip line near $T_{c}$. Another
important result was that the intrinsic magnetic field of the
current and the dynamics of motion of vortices did not affect
significantly the temperature dependence of the resistance of a
phase-slip line. This resistance was governed by $l_{E}$, exactly
as in the case of narrow films containing phase-slip centers.

The authors regard it as their pleasant duty to express their deep
gratitude to V. T. Volkov for ion-beam etching of the structures.

\end{document}